\documentstyle[aps,epsf,prc]{revtex}

\begin{document}
\title{ Associative photoproduction of charmed particles near 
threshold}
\author{Michail P. Rekalo \footnote{ Permanent address:
\it National Science Center KFTI, 310108 Kharkov, Ukraine}
}
\address{Middle East Technical University, 
Physics Department, Ankara 06531, Turkey}
\author{Egle Tomasi-Gustafsson}
\address{\it DAPNIA/SPhN, CEA/Saclay, 91191 Gif-sur-Yvette Cedex, 
France}
\date{\today}

\maketitle
\begin{abstract}
We calculate the cross section and the beam asymmetry for exclusive 
photoproduction of charmed 
particles near threshold ($\gamma+p\to \Lambda_c^+ +\overline{D^0}$), 
in the framework of an 
effective Lagrangian model. We 
discuss the sensitivity of these observables to the magnetic moment of the 
$\Lambda_c$ baryon and of the coupling 
constant in the $p\Lambda_c D$ vertex. We show that exclusive measurements 
allow, in 
principle, to determine the magnetic moments of charmed baryons.
\end{abstract}

The purpose of this paper is to study the associative photoproduction  of 
charmed particles on nucleon.  Among the simplest reactions of open charm 
production,  $\gamma+N\to {\cal B}_c+\overline{D} $, with ${\cal B}_c=\Lambda_c$ 
or 
$\Sigma_c$, 
the 
reaction $\gamma+p\to \Lambda_c^+ +\overline{D^0}$
has the lowest threshold. The experimental data 
about charm photoproduction concern mainly inclusive production of 
$D$$(D^*)$-mesons or 
$\Lambda_c(\Sigma_c)$- baryons at high energies, very far from threshold. The 
lowest photon energy where charm production was observed is 20 GeV, at 
SLAC \cite{Abe84}, with an indirect estimation of the corresponding cross 
section. Up to now no exclusive measurement exists. Photoproduction processes 
involving charmed particles are experimentally 
accessible with a photon beam of energy over 10 GeV. The possibility of a 
systematic study of 
these reactions in the treshold region at future machines 
as ELFE \cite{elfe} or at the Jefferson Laboratory (JLab)  electron accelerator, 
after 
the planned upgrade \cite{gen}, makes this problem very actual.

The photon-gluon fusion,  $\gamma+g\to c+\overline{c}$, is considered as the 
most 
probable mechanism of inclusive photoproduction of charmed particles at high 
energy. In its  different versions (adding, for example, NLO contributions),   
the inclusive spectra of $D^*$ and 
$D$-mesons \cite{Ad99} can be satisfactorily reproduced. However  this 
mechanism 
predicts $\overline{D}/D$ as well 
as $\overline{\Lambda_c}/\Lambda_c$ symmetry, in contradiction with the 
experimental data \cite{tps,Fr96}. Open charm photoproduction can 
induce, at least partly, the observed asymmetry. The standard view of inclusive 
charm photoproduction, based on $\gamma +g-$ fusion, with subsequent 
fragmentation of $c+\overline{c}$ ~into charmed 
particles 
can not be directly applied to the description of the exclusive reactions, in 
any kinematical conditions, and in particular in the threshold region. Also  
other approaches, based on VDM \cite{Fr77} or QCD \cite{No78}, which may 
reproduce the total inclusive cross section, do not give a simple picture of 
exclusive processes.

We will use a formalism based on an effective Lagrangian approach (ELA), 
extrapolating  a  method which is well known in the 
domain of the photoproduction of 
$\pi$- and $K$-mesons.
We will consider the photoproduction of pions, $\gamma +N\to N+\pi$; of strange 
particles, $\gamma+N\to \Lambda+ K$; and of open charm, $\gamma+N\to 
{\cal B}_c +\overline{D}$, as the same class of reactions 
$\gamma+N\to {\cal B}+{\cal P}$ where ${\cal B}$ is a baryon, 
with spin ${\cal J}$ and parity $P$ equal to ${\cal J}^P=1/2^+$, (${\cal B}=N$, 
$\Lambda$, $\Sigma$, $\Lambda_c$, or $\Sigma_c$)  and ${\cal P}$ is the 
corresponding 
pseudoscalar meson (${\cal P}=\pi$, $\eta$, $K$, $D$). In our knowledge, there 
is no principal restriction on the mass of the ${\cal P}$-meson for the 
applicability 
of ELA. Note in this respect, an interesting scaling for the masses of  
pseudoscalar mesons: $m_K/m_\pi\simeq m_D/m_K \simeq 3.6$.

The application of ELA to exclusive charm photoproduction has the following 
advantages:
\begin{itemize}
\item the transparent physical content of the considered mechanisms;
\item limited number of parameters with a definite physical meaning: interaction 
constants and magnetic moments of charmed baryons;
\item theoretical predictions for parameters such as magnetic moments of charmed 
baryons can 
be used in the numerical calculations;
\item the outputs of the model are the absolute value of the differential cross 
section and all polarization observables, in the near threshold region.
\end{itemize}

Such approach has been recently used in the studies of diferent processes 
involving charm particles, as $J/\psi +\pi(\rho)\to D+\overline{D}(D^*)$.
or $J/\psi +N\to \Lambda_C+\overline{D}$ \cite{Ma98,Ha99,Li99,Li00}.
The cross sections for these reactions have been calculated in the ELA 
formalism, 
considering the pole diagrams (with $D$, $D^*$ or $\Lambda_C$ exchanges), in 
relation with the study of the mechanisms responsible for the $J/\psi$ 
suppression in high energy nuclei collisions.

The present analysis should be considered as a starting point in the discussion 
of the possible mechanisms which play the most important role in the threshold 
region \cite{Re1}. The 
predictive power of such approach should be useful in planning the  possible
future measurements of cross sections or polarization observables. 

We found a large sensitivity 
of the  observables to different fundamental parameters: as an example, the 
absolute value of the differential cross section depends mainly on the 
interaction constant in the vertex $N{\cal B}_cD$, while the magnetic moments of 
the 
charmed baryons mostly affect the beam asymmetry. The magnetic moments of 
charmed particles can be calculated in different approaches, from quark models 
to ChPT+HQET considerations,  
\cite{Ch76,Ba91,Sa94,Ba00}. In this respect the 
$\Lambda_c$ is very interesting: the light quarks, here, are in a configuration 
with zero total spin, so the  $\Lambda_c$ magnetic moment is due only to the the 
magnetic moment of the heavy $c-$quark \cite{Ba91}. There is no experimental 
information about the magnetic moments of the charmed 
baryons. The standard methods used for usual hyperons can not be applied here, 
due to the shorter time of life for ${\cal B}_c$. In principle the 
method of bending these particles in crystals can be used \cite{Ch92}. The 
processes 
$\gamma+N\to {\cal B}_c+\overline{D}$ can be considered a parallel and 
independent way.

According to the lines given by a previous 
work \cite{Re1}, which we update and complete, we calculate the differential 
cross section and the excitation function for the process 
$\gamma+p\to \Lambda_c^++\overline{D^0}$. We calculate the beam asymmetry and 
discuss the sensitivity of the different observables to the magnetic moment of 
the $\Lambda_c^+$.

The spin structure for the amplitude of the  processes
$\gamma+N\to {\cal B}_c+\overline{D}$ can be written in a general form
(in the CMS of the considered reaction):
$${\cal F}=i\vec\sigma\cdot\vec e f_1 + 
\vec\sigma\cdot\hat{\vec q}\vec\sigma\cdot\hat{\vec k}\times\vec e f_2+ 
i\vec e \cdot\hat{\vec q}\vec\sigma\cdot\hat{\vec k}f_3
+i\vec\sigma\cdot\hat{\vec q} \vec e \cdot\hat{\vec q}f_4,$$
where $\vec e$ is the photon polarization three-vector, $\hat{\vec k}$ and
$\hat{\vec q}$ are the unit vectors along the three-momentum of $\gamma$ and 
$\overline{D}$.
The differential cross section is given by: 
$$\displaystyle\frac{d\sigma}{d\Omega}=\displaystyle\frac{q}{k}(E_1+m)(E_2+M)
\displaystyle\frac{B+ A\sin^2\vartheta }{128\pi^2s}$$
with 
$$A=|f_3|^2+|f_4|^2+2{\cal R}e f_2f_3^*+2{\cal R}e(f_1+\cos\vartheta 
f_3)f_4^*,$$
$$B=2\left ( |f_1|^2+|f_2|^2-2\cos\vartheta {\cal R}e f_1f_2^*\right ),$$
where $E_1(E_2)$ and $m(M)$ are the energy and the mass of $N({\cal B}_c)$, 
respectively, and $\vartheta$ is the center of mass angle of the $D$-meson 
production (with respect to the direction of the incident photon). In case of a 
linearly polarized photon beam,  the beam asymmetry $\Sigma$ 
can be determinated as:
$$\Sigma=\displaystyle\frac{d\sigma_{\perp}/d\Omega-d\sigma_{\parallel}/d\Omega}
{d\sigma_{\perp}/d\Omega+d\sigma_{\parallel}/d\Omega}= 
\displaystyle\frac{-A\sin^2\vartheta }{B+A\sin^2\vartheta }.$$
The amplitudes $f_i,~i=1-4$, have to be calculated in the 
framework of some model.
We  consider here the pole contributions in the $s-$, $t-$ and $u-$channels 
(Fig. 1), so $f_i=f_{i,s}+f_{i,t}+f_{i,u}$. 

A discussion of the ELA applied to charm particles, with broken SU(4) symmetry, 
can be found in literature \cite{Ma98}. In the framework of this formalism, one 
obtains the following expressions for the matrix element, corresponding to the 
$s-$, $t-$ and $u-$contributions (for the pseudoscalar meson-baryon vertex):
$${\cal M}_s=e\displaystyle\frac {g_{N{\cal B}_cD}}{s-m^2} 
\overline{u}(p_2)\gamma_5(\hat k+\hat p_1+m)\left [ Q_N\hat \epsilon-
\displaystyle\frac {\hat \epsilon \hat k}{2m}\kappa_N\right ]
{u}(p_1),$$
$${\cal M}_u=e\displaystyle\frac {g_{N{\cal B}_cD}}{u-m^2} 
\overline{u}(p_2)
\left ( Q_c \hat \epsilon-
\displaystyle\frac {\hat \epsilon\hat k}{2M}\kappa_C\right  )
(\hat p_2-\hat k+M)\gamma_5 {u}(p_1),$$
$${\cal M}_t=2eQ_D\displaystyle\frac {g_{N{\cal B}_cD}}{t-M_D^2}
\hat \epsilon\cdot \hat q \overline{u}(p_2)\gamma_5 {u}(p_1),$$
where $k$, $q$, $p_1$ and $p_2$ are the four momenta of $\gamma$, $D$, $N$ and 
$\Lambda_c$, 
$\epsilon $ $(\epsilon \cdot k=0)$ is the four vector of the photon 
polarization, $s=(k+p_1)^2$, $u=(k-p_2)^2$, $t=(k-q)^2$ are the Mandelstam 
variables, $m$ ($Q_N$), $M$ ($Q_c$) and $M_D$ ($Q_D$) are  the masses (electric 
charges) of the nucleon,
the charmed baryon and the $D$-meson, respectively, $g_{N{\cal B}_cD}$ is 
the coupling constant for the vertex of nucleon-charmed baryon-$D$ meson 
interaction; $\kappa_N$ and $\kappa_c$ are 
the nucleon and charmed baryon anomalous magnetic moment ($\kappa_N=1.79$ 
(-1.91) for $p$ ($n$)).

From these formulas we can derive the following expressions
for the scalar amplitudes $f_i$.

\noindent\underline{\bf s-channel}:
$$f_{1,s}=e\displaystyle\frac{g_{N{\cal B}_cD}}{W+m}
\left [ 
Q_N-(W-m)\displaystyle\frac{\kappa_N}{2m}\right ],$$
$$f_{2,s}=e\displaystyle\frac{g_{N{\cal B}_cD}}{W+m}\left [ 
-Q_N-(W+m)\displaystyle\frac{\kappa_N}{2m}\right ]
\displaystyle\frac{|\vec q|}{E_2+M},$$
$$f_{3,s}=f_{4,s}=0,~\displaystyle\frac{e^2}{4\pi}
=\alpha\simeq\displaystyle\frac{1}{137}, $$
where $W=\sqrt{s}$, is the total energy, related to the $\gamma$-energy 
in the laboratory system by $s=m^2+2E_\gamma m$.

\noindent\underline{\bf u-channel}:
$$f_{1,u}=e\displaystyle\frac{g_{N{\cal B}_cD }}
{u-M^2}
\left \{ Q_c (W-M)-\displaystyle\frac{\kappa_c}{2M}
\left [ t-M_D^2+(W-m)(W+m-2M)\right ]\right \},$$
$$f_{2,u}=- e\displaystyle\frac{g_{N{\cal B}_cD}}
{u-M^2}
\displaystyle\frac{|\vec q|}{E_2+M}
\left \{ Q_c (W+M)+\displaystyle\frac{\kappa_c}{2M}
\left [ t-M_D^2+(W+m)(W-m+2M)\right ]\right \}\displaystyle\frac{W-m}{W+m},$$
$$f_{3,u}= e\displaystyle\frac{g_{N{\cal B}_cD}}{u-M^2}|\vec 
q|\displaystyle\frac{W-m}{W+m}
\left [ 2Q_c+\kappa_c\displaystyle\frac{W+m}{M}\right ],$$
$$f_{4,u}=\displaystyle\frac{g_{N{\cal B}_cD e}}{u-M^2}(E_2-M)\left [- 
2Q_c+\kappa_c\displaystyle\frac{W-m}{M}\right ],$$

\noindent\underline{\bf t-channel}:
$$f_{1,t}=f_{2,t}=0,$$
$$f_{3,t}=-2 eQ_D\displaystyle\frac{g_{N{\cal B}_cD}}{t-M^2_D}|\vec 
q|\displaystyle\frac{W-m}{W+m},$$
$$f_{4,u}=2 eQ_D\displaystyle\frac{g_{N{\cal B}_cD}}{t-M^2_D}(E_2-M).$$

The  angular distribution and 
the 
beam asymmetry for the process $\gamma+p\to \Lambda_c^++\overline{D^0}$ are 
reported on Fig. 
2 as functions of $\vartheta$. The full line represents the sum of all 
contributions, the  
Born $s$-channel term is given by the dashed line and the  Born $u$-channel term 
by the 
dotted line. The $t-$ channel diagram does not play any role 
for this reaction in the present model, as $Q_D=0$.
Even in the case when the contribution of one diagram is negligible its 
interference with the 
other terms can largely affect the total result: 
a large  effect of the $s$-$u-$ interference may appear in the differential 
cross 
section (Fig. 2).
  
The plotted differential  cross section  is divided 
by $g_{ND\Lambda_c}^2$, for which we do not have (theoretical 
or experimental) precise indications. 

The behavior of the $\Sigma-$asymmetry results from the difference in the spin 
structure of the $s$ and $u-$ channel contributions to the matrix element for 
$\gamma+p\to\Lambda_c+\overline{D_0}$. The $s-$ channel diagram (mainly 
characterized by s-wave $\Lambda_c D$-production with a small off-mass shell 
admixture of p-wave) can not induce nonzero $\Sigma-$asymmetry, contrary to the 
$u-$channel contribution, which contains the majority of the multipole 
amplitudes, even in the threshold region, and is responsible for the  
positive value of $\Sigma$, in the whole angular domain. However the final 
result is essentially driven by the $s-u$-interference, which is negative, 
compensating the positive $u-$channel  contribution.

In order to illustrate the sensitivity of this reaction to the $\Lambda_c$ 
magnetic moments, in Fig. 3 we show, as a function of $\mu_{\Lambda_c}$,
the dependence of the , taken at $\vartheta=\pi /2$ and of the ratio 
of the differential cross section for $\vartheta=\pi$ and $\vartheta=0$ 
(backward-forward asymmetry). Both these quantities show a characteristic 
dependence on $\mu_{\Lambda_c}$, in the region $\mu_{\Lambda_c}\le 1$, being 
almost independent on  $\mu_{\Lambda_c}$ for $\mu_{\Lambda_c}\ge 1$. The 
asymmetry $\Sigma(\pi/2)$ changes sign, for $\mu_{\Lambda_c}\simeq 1$.

The sensitivity of the absolute cross section to $\mu_{\Lambda_c}$ appears 
explicitely at threshold, where 
only the amplitude $f_1$ is present and can be written as:
$$f_1\simeq 1- \displaystyle\frac{W-m}{2M_D}\left (\mu_{\Lambda_c} 
-1+\kappa_p\frac{M}{m}\right )\propto 1+0.46 ~\mu_{\Lambda_c}.
$$

The excitation functions for all the considered reactions, with magnetic moments 
as reported in the Table, are shown in Fig. 4: $\gamma+p\to \Lambda_c^+ 
+\overline{D^0}$ (thick solid line),
$\gamma+p\to\Sigma_c^{++}+D^-$ (dashed line), 
$\gamma+p\to \Sigma_c^+  + \overline{D^0}$ (dotted line), 
$\gamma+n \to \Lambda_c^+ +D^-$  (dot-dashed line), 
$\gamma+n \to \Sigma_c^+  + D^-$ (solid line), 
$\gamma+n \to \Sigma_c^0  + \overline{D^0}$ (thick dot-dashed line). The 
calculation for 
the reaction $\gamma+p\to \Lambda_c^+ +\overline{D^0}$, using the 
magnetic moment suggested by \cite{Sa94} is reported as a thick dashed line.

These calculations reproduce satisfactorily the measurements at high energy (on 
proton target) but 
overestimate
the measurement at the lowest energy, from \cite{Abe84}.  The interest of such a 
comparison with the  
experimental is to give an estimation of the upper limit of the constant 
$g_{N{\cal B}_cD}$, here taken as unity. 

In the present model, the cross section for the process 
$\gamma+n\to\Sigma_c^0+D^0$ is larger  when compared with the other processes, 
due to the fact
that the magnetic moments of $n$ and $\Sigma_C^0$ contribute coherently in the 
$s-$ and $u-$ channel.

Note that the possible T-odd 
polarization observables have to be 
identically zero, in the framework of the considered model, in any kinematical 
conditions, for any reaction $\gamma +N\to {\cal B}_c^+ +\overline{D}$.
     
In conclusion, we have calculated the differential cross section and the beam 
asymmetry for the exclusive photoproduction of charmed particles near 
threshold, $\gamma+N \to {\cal B}_c+\overline{D}$. 
We have studied the effects of the magnetic 
moments of charmed baryons and of the coupling constant for the $N{\cal B}_cD$ 
vertex 
and discussed the possibility to determine these quantities, which are 
not known experimentally.
We have shown that the absolute value of the cross section depends mainly on the 
interaction constant in the vertex $N{\cal B}_cD$. On the other hand, the 
angular 
dependence and 
even the sign of the beam asymmetry is largely affected by the value of the 
magnetic moment of the charmed baryon.


\begin{table*}
\begin{tabular}{|c|c|c|c|c|c|}
Reaction &Threshold [GeV]& $\mu_{\Lambda_c,\Sigma_c}~[\mu_N]$&
$\kappa_c~[\mu_N]$&$M_{{\cal B}_c}~[GeV]$&$M_D~[GeV]$\\
\hline\hline
$\gamma+p\to \Lambda_c^+ +\overline{D^0}$   &8.7064  & 1.86 & 0.86 & 2.2849 
&1.8646 \\
\hline\hline
$\gamma+p\to \Sigma_c^{++}+D^-$ &9.4856  & 1.86 &-0.14 & 2.4528  &1.8693 \\
\hline\hline
$\gamma+p\to \Sigma_c^+  + \overline{D^0}$ &9.4677  & 0     & -1 & 2.4536 & 
1.8646 \\
\hline\hline
$\gamma+n \to \Lambda_c^+ +D^-$  &8.7139  & 1.86 & 0.86 & 2.2849 &1.8693 \\
\hline\hline
$\gamma+n \to \Sigma_c^+  + D^-$ &9.4750 &  0    & -1    & 2.4536  &1.8693 \\
\hline\hline
$\gamma+n \to \Sigma_c^0  + \overline{D^0}$ &9.4469 &-1.86 &-1.86 & 2.4522 
&1.8646 \\
\end{tabular}
\caption{ Table of reactions and constants.}
\label{tab1}
\end{table*}

\begin{figure}
\caption{Feynman diagrams calculated for the process 
$\gamma+N\to 
{\cal B}_c+\overline{D}$: (a)- $s$-channel,  (b)- $u$-channel, (c)- 
$t$-channel.
}
\end{figure}

\begin{figure}
\caption{Differential cross 
section and beam asymmetry as a function of the $D$-meson cms angle, 
$\vartheta$:  full 
calculation  (solid line), Born $s$-channel (dashed line), and Born $u$-channel 
(dotted line).}
\end{figure}
\begin{figure}
\caption{Asymmetry calculated at $\vartheta=\pi /2$ (top)
and ratio of the cross section for $\vartheta =\pi$ and $\vartheta=0$ 
(bottom) as functions of the $\Lambda_c$ magnetic moment.}
\end{figure}
\begin{figure}
\caption{Sample of existing experimental data from 
\protect\cite{Abe84} (circle), \protect\cite{na14} (square), 
\protect\cite{emc83} (triangles),
\protect\cite{Wa43} (reversed triangle), \protect\cite{tps} (open circle), 
 \protect\cite{PhEC87} (open square).
The different lines show the calculations for the different reactions (see text 
and Table), asssuming the interaction constant  $g_{N{\cal B}_cD}=1$.}
\end{figure}


\begin{references}
\bibitem{Abe84} K. Abe  et al., Phys. Rev. D 30 (1984) 1.
\bibitem{elfe} J. Arvieux and E. de Sanctis  The ELFE Project, (Ed. 
Compositori, Bologna, 1992).
\bibitem{gen} See the website http://www.jlab.org/~gen/charm/.

\bibitem{Ad99} C. Adloff  et al., H1 Coll., Nucl. Phys. B 545 (1999) 21;\\
J. Breitwag  et al., ZEUS Coll., Eur. Phys. Journ. C 6(1999) 67; C 12 (2000) 35.
\bibitem{tps} J. C. Anjos  et al., Tagged Photon Spectrometer Coll., Phys. 
Rev. Lett.  62  (1989) 513.
\bibitem{Fr96} P. L. Frabetti  et al., E687 Coll., Phys. Lett. B 370  
(1996) 222.
\bibitem{Fr77}  H. Fritsch, Phys. Lett. B 67, (1977) 217;\\ 
               H.  Fritsch and K. H. Streng, Phys. Lett. B 78 (1978) 447.
\bibitem{No78} V. A. Novikov  et al., Nucl. Phys. B 136 (1978) 12.
\bibitem{Ma98} S. G. Matinyan and B. Muller, Phys. Rev. C 58 (1998) 2994.
\bibitem{Ha99} K. L. Haglin, nucl-th/9907034 (1999).
\bibitem{Li00}Z. Liu and  C. M. Ko nucl-th/9912046 (1999)
\bibitem{Li99} Z. Liu, C. M. Ko and B. Zhang, Phys. Rev. C 61 (2000) 024904.

\bibitem{Re1} M. P. Rekalo, Ukr. Fiz. Journ. 22 (1977) 1602. 
\bibitem{Ch76} A. L. Choudhury and V. Joshi, Phys. Rev. {\bf D13}, 3120 (1976).
\bibitem{Ba91} D.-P. Min, Proc.  Int. Workshop 'Baryon Spectroscopy 
and the structure of the Nucleon (Saclay, 23-25 Sept. 1991) p. 138. 
\bibitem{Sa94} M. J. Savage, Phys. Lett. B 326, (1994) 303.
\bibitem{Ba00} M. C. Banuls   et al., Phys. Rev. Lett.  D 61 (2000) 074007.
\bibitem{Ch92} D. Chen  et al., E761 Coll., Phys. Rev. Lett. 69 (1992) 3286.
\bibitem{na14} M . Alvarez  et al.,  NA 14/2 Coll. Z. Phys. C 60 
(1993) 53.
\bibitem{emc83} J.J. Aubert  et al., EMC Coll. Nucl. Phys B 213 (1983) 31. 
\bibitem{Wa43}  D. Aston  et al., WA4 Coll.  Phys. Lett. B 94 (1980) 113 .
\bibitem{PhEC87} M. J. Adamovich  et al., Photon Emulsion Coll., Phys. 
Lett.  B 187 (1987) 437.

\end{references}
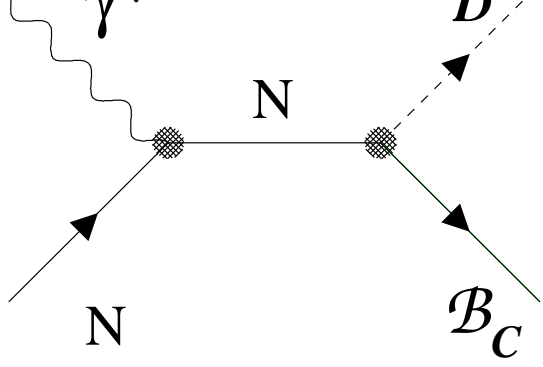
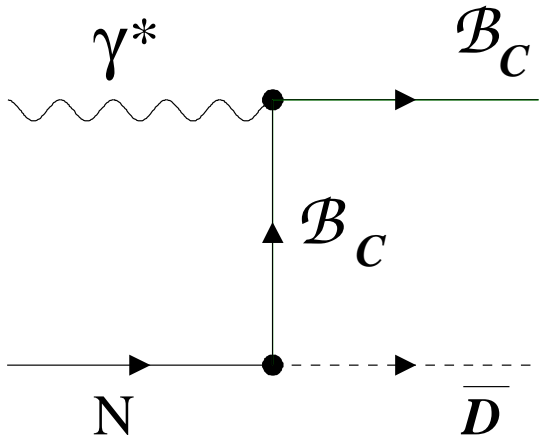
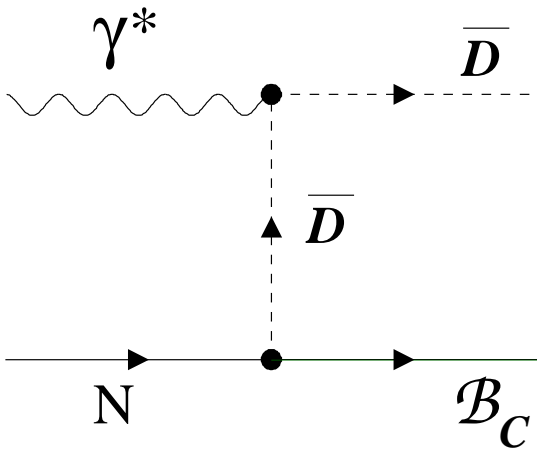
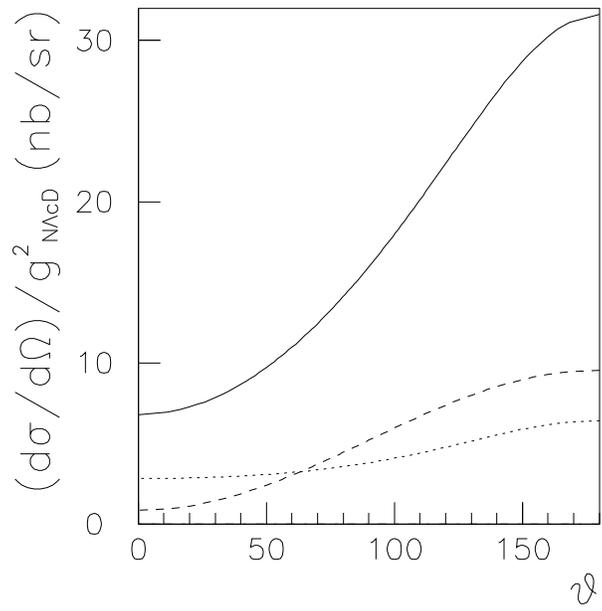
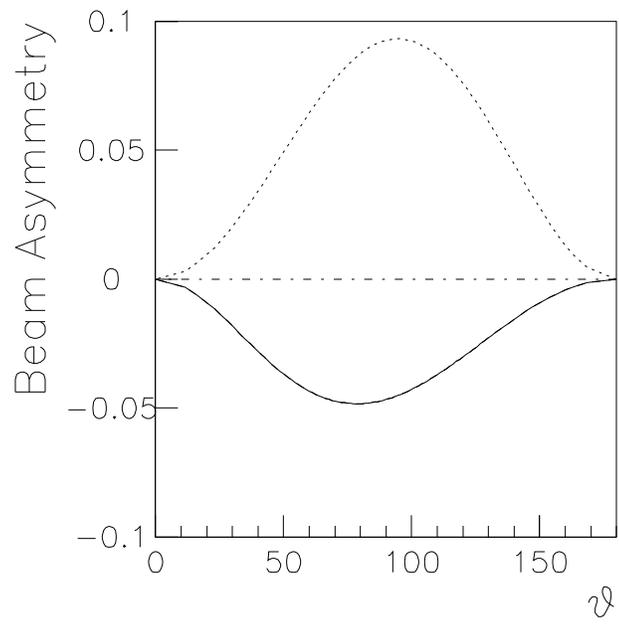
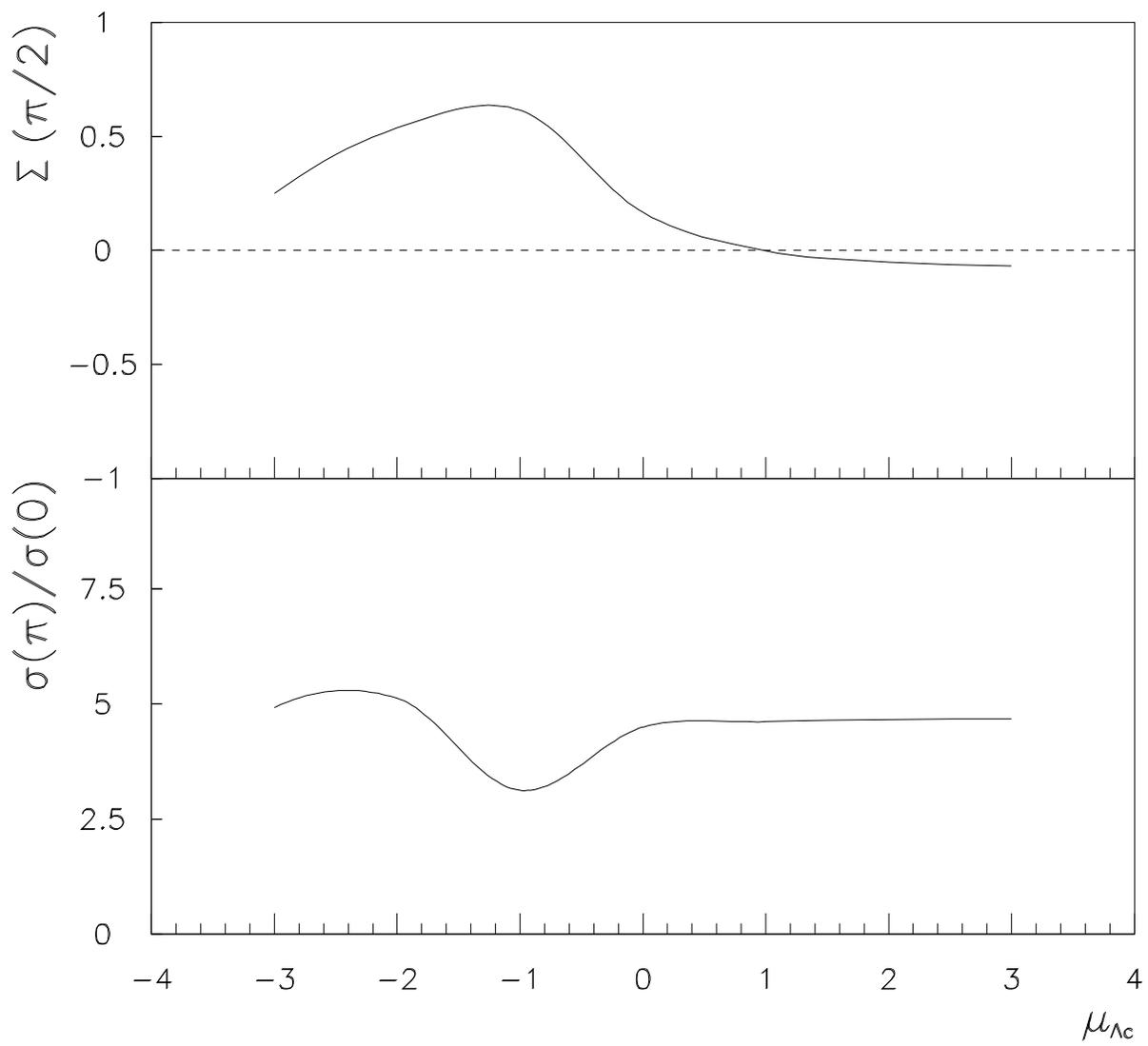
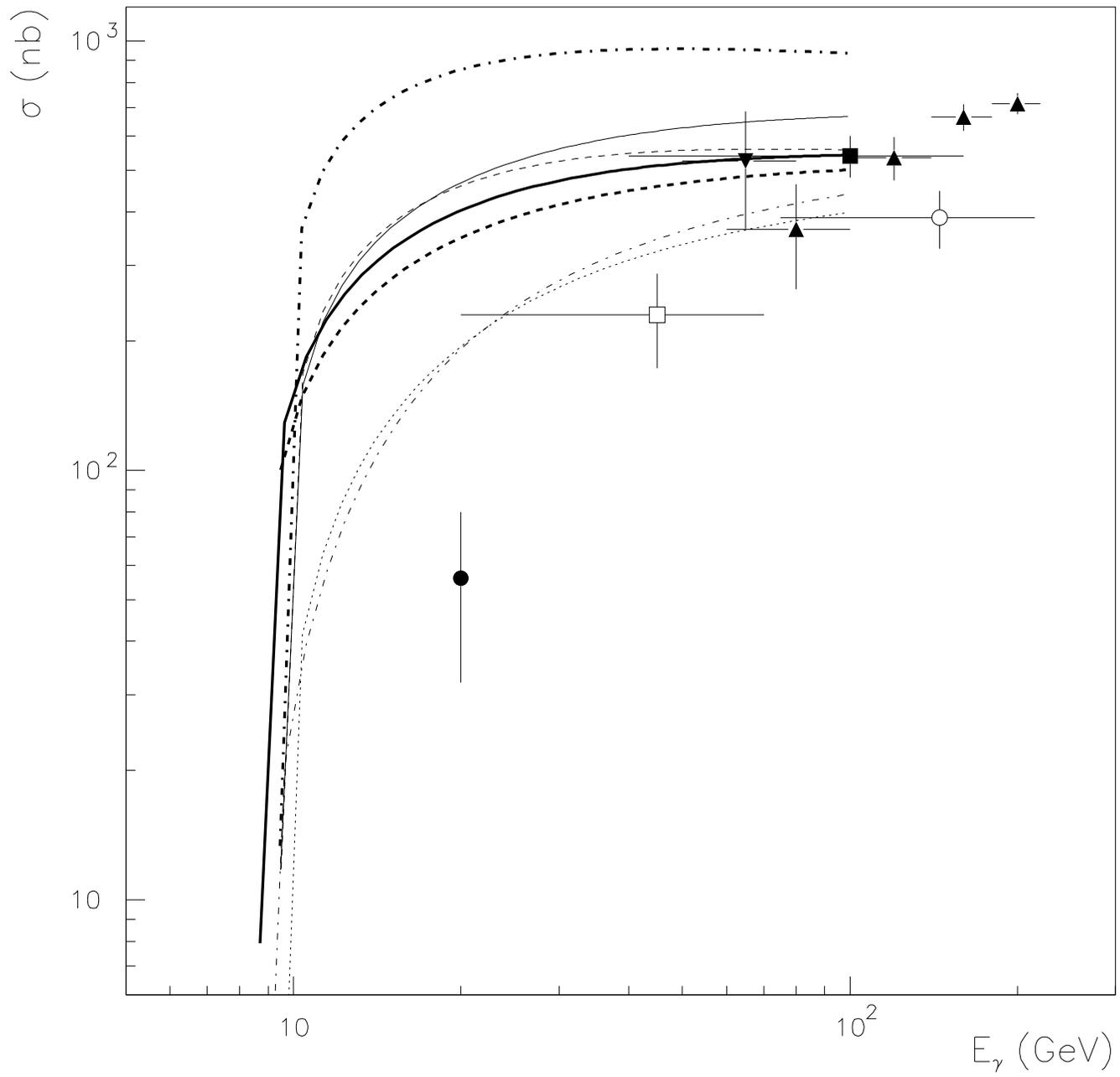
\end{document}